\documentclass[12pt]{article}

\usepackage{amssymb}
\usepackage{amsmath}
\usepackage{amscd}
\usepackage{latexsym}
\usepackage{graphicx}

\usepackage{cite}

\newcommand{\be}{\begin{equation}}
\newcommand{\ee}{\end{equation}}
\newcommand{\Dlt}{\Delta}

\newcommand{\prt}{\partial}

\newcommand{\bt}{\beta}
\newcommand{\vp}{\varphi}
\newcommand{\ep}{\varepsilon}
\newcommand{\al}{\alpha}
\newcommand{\ra}{\rightarrow}

\newcommand{\gm}{\gamma}

\newcommand{\Gm}{\Gamma}

\newcommand{\lbd}{\lambda}

\begin{document}

\begin{center}
{\Large {\bf Self-similar renormalization for nonlinear problems} \\ [5mm]

V.I. Yukalov$^{1,2}$ and E.P. Yukalova$^3$} \\ [3mm]

{\it $^1$Bogolubov Laboratory of Theoretical Physics, \\
Joint Institute for Nuclear Research, Dubna 141980, Russia \\ [2mm]

$^2$Instituto de Fisica de S\~ao Carlos, Universidade de S\~ao Paulo, \\
CP 369, S\~ao Carlos 13560-970, S\~ao Paulo, Brazil \\ [3mm]

$^3$Laboratory of Information Technologies, \\
Joint Institute for Nuclear Research, Dubna 141980, Russia }  

{\bf E-mail}: yukalov@theor.jinr.ru

\end{center}

\vskip 2cm

\begin{abstract}
A new method, called the method of self-similar approximants, and its recent 
developments are described. The method is based on the ideas of renormalization 
group theory and optimal control theory. It allows for the effective extrapolation 
of asymptotic series in powers of small variables to the finite and even to infinite 
variables. The approach is proved to be regular. It is illustrated by several 
examples demonstrating good agreement with numerical calculations. The method is 
shown to provide accurate approximate solutions to complex nonlinear problems. In 
some cases, the method allows for the reconstruction of exact solutions on the basis 
of rather short perturbative series.
\end{abstract}

\newpage

\section{Introduction}

Divergent series are ubiquitous in physics and applied mathematics, when problems are 
solved employing a kind of perturbation theory in powers of some parameter assumed 
to be small. Such series, as a rule, are asymptotic and diverge for finite values of the 
expansion parameter. However, in practical applications one usually needs to know the 
observable quantities at finite values of the parameter, which requires to invent methods 
of defining effective limits of the series, which would be valid for finite values of the
expansion variable. Sometimes, one even needs to find the observable quantities in the
limit of infinitely large variables. There exist the methods of extrapolation of the 
small-variable expansions to finite values of the latter, such as the methods of
Pad\'{e} approximants \cite{Baker_1}, Borel summation \cite{Glimm_2,Hardy_3}, Shanks 
sequence transformations \cite{Weniger_4} and others. However, the most difficult problem 
of extrapolating series from the asymptotically small variables to the limit of the 
infinite variables has been illusive. 

In the present communication, we report on an original method allowing for the 
extrapolation of series of asymptotically small variables to arbitrary values of the 
variables, including their asymptotically large values. The method is based on self-similar
approximation theory \cite{Yukalov_5,Yukalov_6,Yukalov_7,Yukalov_8, Yukalov_9} (see also 
recent review articles \cite{Yukalov_10,Yukalov_11}). We formulate the basic points of the 
self-similar approximation theory and give examples of its application for finding 
strong-coupling limits of several problems.

\section{Basic points of self-similar approximation theory}

Suppose we are looking for an observable quantity $f(x)$ expressed in the form of an 
expansion in powers of a variable $x \ra 0$. The function $f(x)$ is real-valued, and the 
variable $x$ is also real-valued. The $k$-th order expansion reads as
\be
\label{1}
f_k(x) \; = \; \sum_{n=0}^k a_n x^n \qquad ( x \ra 0) \;   .
\ee
In general, the series for $f_k(x)$ can be divergent for any finite $x$. Our aim is to
extrapolate this series to arbitrary values of $x$, so that the extrapolated function 
would be valid for all finite $x$, as well a for $x \ra \infty$. The main pillows of the
self-similar approximation theory are as follows \cite{Yukalov_10,Yukalov_11}. 

1. {\it Implantation of control parameters}. In order to induce convergence for the given
sequence, it is necessary to reorganize the latter by introducing control parameters,
so that the terms $f_k(x)$ are transformed into $F_k(x,u)$, where $u$ is a set of control 
parameters. These parameters can be implanted in different ways, through initial conditions,
through the change of variables, or by sequence transformation.

2. {\it Determination of control functions}. The implanted control parameters $u$ are to 
be converted into control functions $u_k(x)$ to make the sequence of the terms 
$F_k(x,u_k(x))$ convergent. Control functions are defined by minimizing a cost functional
$C[u]$. For instance, the cost functional can be taken in the form
\be
\label{2}
 C[\; u \; ] \; = \; \lbd |\; F_{k+1}(x,u) - F_k(x,u) \; |^2 +
( 1 -\lbd) \left| \; \frac{\prt F_k(x,u)}{\prt u} \; \right|^2 \;  .
\ee
The other way is to define $u_k$ from training conditions.
 
3. {\it Construction of approximation cascade}. The sequence of the approximants 
$\{F_k(x,u_k)\}$ is represented by a bijective sequence $\{y_k(f)\}$ that is treated 
as a trajectory of a dynamical system, approximation cascade, where the approximation 
order $k$ plays the role of discrete time and $f$ is an initial condition. The bijective 
terms $F_k$ and $y_k$ are images of each other. The transition from one approximation 
term to another is treated as the motion in the approximation space. The dynamics of 
the cascade in the vicinity of a fixed point enjoys the property of self-similarity
\be
\label{3}
y_{k+p}(f) \; = \; y_k( y_p(f) ) \;   .
\ee

4. {\it Embedding cascade into flow}. The approximation cascade, characterizing the 
motion in discrete time, can be embedded into a flow that is a dynamical system in
continuous time. The trajectory of the approximation flow passes through all points of 
the cascade trajectory, with the property of self-similarity being preserved,
\be
\label{4}
 y(t+t',f) \; = \; y( t,y(t',f) ) \;  .
\ee
   
5. {\it Retrieving fixed points}. The self-similar relation (\ref{4}) can be represented
in the form of the Lie differential equation
\be
\label{5}
\frac{\prt}{\prt t}\; y(t,f) \; = \; v(y)
\ee
that can be integrated yielding the evolution integral
\be
\label{6}
 \int_{y_k}^{y_k^*} \frac{dy}{v_k(y)} \; = \; t_k \;  ,
\ee
where $v_k$ is the flow velocity and $t_k$ is effective time for reaching the fixed 
point $y^*_k$. This fixed point $y^*_k$ is bijective to the effective limit of the 
approximation sequence, $f_k^*(x)$, called self-similar approximant of order $k$.   
    
As we see, the basics of self-similar approximation theory employ the ideas of 
renormalization-group theory \cite{Kleinert_12}, dynamical theory \cite{Katok_13}, 
optimal control theory \cite{Lee_14}, and machine learning \cite{Murphy_15,Alpaydin_16}. 
          
Control parameters can be incorporated into the sequence of terms (\ref{1}) by means 
of the fractal transform
\be
\label{7}
 F_k(x,\{ n_j\} ) \; = \; \prod_{j=1}^k x^{-n_j} ( 1 + b_j x) \; ,
\ee
where $n_j$ and $b_j$ are control parameters. Fixed points are given by $y^*_k$ whose
images are the self-similar factor approximants \cite{Yukalov_17,Gluzman_18}
\be
\label{8}
f_k^*(x) \; = \; \prod_{j=1}^{N_k} ( 1 + A_j x)^{n_j} \;   ,
\ee
where the control parameters now are $A_j$ and $n_j$ and
\begin{eqnarray}
\nonumber
N_k \; = \; \left\{ \begin{array}{ll}
k/2 \; , ~ & ~ k = 2,4,\ldots \\
(k+1)/2 \; , ~ & ~ k = 3,5,\ldots
\end{array} \right. 
\end{eqnarray}

The control parameters are defined from the training conditions
\be
\label{9}
 f_k^*(x) - f_k(x) \; \simeq \; 0 \qquad ( x \ra 0 ) \;  ,
\ee
which yield the conditions
\be
\label{10}
 \lim_{x\ra 0} \; \frac{1}{n!} \; \frac{d^n}{dx^n} \; f_k^*(x) \; = \; a_n \;  .
\ee

Also, it is possible to use the Borel transform
\be
\label{11}
 B_k(x) \; = \; \sum_{n=0}^k \frac{a_n x^n}{\Gm(n+1+u)} \;  ,
\ee
with $u$ being a control parameter, extrapolate (\ref{11}) by means of the 
factor-approximants
\be
\label{12}
 B_k^*(x) \; = \; \frac{a_0}{\Gm(1+u)} \prod_{j=1}^{N_k} ( 1 + A_j x)^{n_j} \;  ,
\ee
and accomplish the inverse transformation resulting in the factor approximant
\be
\label{13}
 f_k^*(x) \; = \; \int_0^\infty e^{-t} \; t^u \; B_k^*(xt) \; dt \;  .
\ee
If we are interested in the large-variable (strong-coupling) limit, then we obtain
\be
\label{14}   
 f_k^*(x) \; \simeq \; C_k x^{\nu_k} \;  ,
\ee
where the amplitude and exponent are
\be
\label{15}
C_k \; = \; \frac{\Gm(1+u_k+\nu_k)}{\Gm(1+u_k)} \prod_{j=1}^{N_k} A_j^{n_j} \; ,
\qquad
\nu_k \; = \; \sum_{j=1}^{N_k} n_j \;   .
\ee

Thus, given an expansion (\ref{1}) for an asymptotically small parameter or variable 
$x$, its extrapolation to the whole region of the variable $x \in [0,\infty)$ can be 
done either by directly summing this expansion by self-similar factor approximants 
(\ref{8}) or, by involving the Borel transformation, summing the Borel transformed 
series using the factor approximants, and then defining the integral (\ref{13}). The 
method is shown to be regular \cite{Yukalov}.

The most difficult, although interesting, region of the variable $x$ is the limit of 
large $x \ra \infty$, which we shall pay a special attention to. For defining the 
power-law at large $x$, it is also possible to employ the self-similar factor approximants 
for the diff-log transformed series. The details can be found in the reviews 
\cite{Yukalov_10,Yukalov_11} and recent papers \cite{Yukalov_19,Yukalov_20}.

\section{Examples of application}

The method has been applied to a variety of problems, some of which are mentioned here.

{\bf Exact restoration of functions}. Sometimes the extrapolation of asymptotic 
small-variable expansions allows for recovering the related functions exactly. For 
instance, it is straightforward to check that constructing the self-similar factor 
approximants for the expansion $1 + x + x^2/2$, one exactly restores the exponential 
function $e^x$ in all orders $k \geq 2$. 

Another example is given by extrapolating the weak-coupling series  
\be
\label{16}
\bt_k(g) \; = \; - \; \frac{3g^2 N_c}{16\pi^2} \sum_{n=0}^k b_n g^{2n} \qquad
(g\ra 0 ) \;   ,
\ee
related to the beta-function of supersymmetric Yang-Mills theory, where 
$b_n = (N_c/8 \pi^2)^n$, with $N_c$ being the number of colors. In all orders $k \geq 2$,
we get the beta function
\be
\label{17}
\bt_k^*(g) \; = \; - \; \frac{3g^2 N_c}{16\pi^2} \; \left( 1 - \; 
\frac{N_c}{8\pi^2}\; g^2 \right)^{-1} \;  ,
\ee
coinciding with its exact form \cite{Novikov_22,Novikov_23,Shifman_24,Arkani_25,Arkani_26}.  

{\bf Exact solutions of differential equations}. Solutions to nonlinear differential 
equations can be derived by, first, finding an approximate solution in the form of a 
series in powers of a variable and then by extrapolating the latter by means of 
self-similar factor approximants. It turns out, that in some cases factor approximants 
reconstruct exact solutions of these equations \cite{Yukalova_27}. Consider, for example, 
the singular equation
\be
\label{18}
 \frac{\ep}{2} \; \frac{d^2\vp}{d x^2} + \vp - \vp^3 \; = \; 0 \;  , 
\ee
where $\varepsilon$ is a positive parameter and the boundary conditions are 
$\varphi(\pm \infty) = \pm 1$. Using the change of the variable 
$x = (\sqrt{\varepsilon}/2) \ln z$, looking for a solution in the form of the 
expansion in powers of $z$, and applying the factor approximants, we get the exact
kink-soliton solution
\be
\label{19}
\vp_k^*(x) \; = \; \tanh\left( \frac{x}{\sqrt{\ep}} \right) \qquad
( k \geq 4) \;   .
\ee

Another example is provided by the singular equation
\be
\label{20}
  \frac{\ep}{2} \; \frac{d^2\vp}{d x^2} - \vp + \vp^3 \; = \; 0 \;   ,   
\ee
with the boundary condition $\varphi(\pm \infty) = 0$. Employing the same procedure, 
as in the previous case, we obtain the self-similar factor approximant
\be
\label{21}
\vp_k^*(x) \; = \; \sqrt{2} \; {\rm sech}\left( \sqrt{\frac{2}{\ep}} \; x \right) \qquad
( k \geq 3) \;
\ee
recovering the exact bell-soliton solution. 

\vskip 2mm
{\bf Critical temperature in $N$-component field theory}. The critical temperature 
$T_c(\gamma)$ in the $N$ component field theory can be represented as a function of 
the coupling parameter $\gamma = \rho^{1/3} a_s$, where $\rho$ is average density and 
$a_s$ is scattering length. One studies the shift of the critical temperature when 
switching on the coupling parameter,
\be
\label{22}
 \frac{\Dlt T_c}{T_c} \; \equiv \; \frac{T_c(\gm)-T_0}{T_0} \; ,
\qquad
T_0\; \equiv \; \frac{2\pi}{m} \; \left[ \; 
\frac{\rho}{\zeta(3/2)} \; \right]^{2/3} \;  .
\ee
This shift, at small $\gamma$, is known to be linear \cite{Arnold_28}, so that
$\Delta T_c/ T_c \simeq C \gamma$, as $\gamma \ra 0$. The coefficient $C$ can be 
expressed \cite{Kastening_29,Kastening_30,Kastening_31} as a function of a variable 
$x$ tending to infinity, such that $C = \lim C(x)$ at $x \ra \infty$. However, in 
practice, the function $C(x)$ can be found only as an asymptotic expansion at small 
$x \ra 0$. Extrapolating this expansion by means of the factor approximants, we find 
\cite{Yukalov_32,Yukalov_33} the values: $C = 0.77$ for $N = 0$; $C = 1.06$ for $N = 1$; 
$C = 1.29$ for $N = 2$; $C = 1.46$ for $N = 3$; and $C = 1.60$ for $N = 4$. These values 
are in perfect agreement with Monte Carlo simulations 
\cite{Prokofiev_34,Arnold_35,Arnold_36,Sun_37}.  

\vskip 2mm
{\bf Critical exponents}. Critical exponents can be presented as $\varepsilon$-expansions 
in powers of $\varepsilon = 4 - d$. These expansions can be extrapolated by using the
self-similar factor approximants and then substituting $\varepsilon = 1$. The results are
found \cite{Yukalov,Yukalov_38} to be in very good agreement with Monte Carlo 
simulations 
\cite{Hasenbusch_40,Hasenbusch_41,Hasenbusch_42,Clisby_43,Clisby_44,Hasenbusch_45}.
 
\vskip 2mm
{\bf Partition function for zero-dimensional field theory}. The partition function 
\be
\label{23}
Z(g) \; = \; \frac{1}{\sqrt{\pi}} \int_{-\infty}^\infty e^{-\vp^2 - g \vp^4}\; d\vp
\qquad ( g \geq 0)
\ee
can be represented as an expansion in powers of the coupling parameter $g \ra 0$. 
This expansion can be extrapolated by using self-similar factor approximants 
\cite{Yukalov_46,Yukalov_47}. The effective sum, found by means of the self-similar 
Borel summation, yields \cite{Yukalov_19,Yukalov_20} the strong-coupling limit
\be
\label{24}
Z_{11}^*(g) \; \simeq \; 0.978 g^{-0.243} \qquad ( g\ra \infty) \;   ,
\ee
as compared with the exact limit
\be
\label{25}
Z(g) \; \simeq \; 1.023 g^{-0.25} \qquad ( g\ra \infty) \; .
\ee
    
\vskip 2mm
{\bf Ground-state energy of anharmonic oscillator}. The ground-state energy of 
one-dimensional quartic oscillator can be obtained as an expansion in powers of the
coupling parameter $g \ra 0$. Employing the self-similar Borel summation, the  
energy can be extrapolated \cite{Yukalov_20} to all values of $g$, including the 
strong-coupling limit
\be
\label{26}
E_{11}^*(g) \; \simeq \; 0.695 g^{0.324} \qquad ( g\ra \infty) \; ,
\ee 
that is very close to the exact expression
\be
\label{27}
E(g) \; \simeq \; 0.668 g^{1/3} \qquad ( g\ra \infty) \;  .
\ee

\vskip 2mm
{\bf Gell-Mann--Low function of $N$-component $\varphi^4$ theory}. This function is 
represented \cite{Kompaniets_48,Schnets_49} as an expansion in powers of the coupling
parameter $g \ra 0$. The extrapolation to arbitrary values of $g$ can be done by using
self-similar factor approximants \cite{Yukalov_19,Yukalov_20}. Depending on the number 
of components $N$, we have
$$
\bt_5^*(g) \; \simeq \; 1.698 g^{1.764} \qquad ( N = 0) \; ,
$$
$$
\bt_6^*(g) \; \simeq \; 1.857 g^{1.750} \qquad ( N = 1) \; ,
$$
$$
\bt_5^*(g) \; \simeq \; 2.017 g^{1.735} \qquad ( N = 2) \; ,
$$
$$
\bt_5^*(g) \; \simeq \; 2.178 g^{1.719} \qquad ( N = 3) \; ,
$$
\be
\label{28}
\bt_5^*(g) \; \simeq \; 2.340 g^{1.702} \qquad ( N = 4) \;   .
\ee
  
\vskip 2mm
{\bf Gell-Mann--Low function in quantum electrodynamics}. The Gell-Mann--Low function 
in quantum electrodynamics, written as an expansion in powers of the coupling parameter 
$\alpha \ra 0$ \cite{Kataev_50}, can be extrapolated by self-similar factor approximants
\cite{Yukalov_19,Yukalov_20} giving
\be
\label{29}
\bt_4^*(\al) \; \simeq \; 0.476 \; \left( \frac{\al}{\pi}\right)^{2.096}
\qquad ( \al \ra \infty) \; .
\ee

{\bf Gell-Mann--Low function in quantum chromodynamics}. In quantum chromodynamics, the 
Gell-Mann--Low function can also be derived as an expansion in powers of the quark-gluon
coupling parameter \cite{Luthe_51,Baikov_52,Herzog_53}. By self-similar factor 
approximants, it can be extrapolated to arbitrary values of $\alpha_s$. In the 
strong-coupling limit, this gives
\be
\label{30}
\bt_2^*(\al_s) \; \simeq \; - 0.366 \al_s^{2.598} \; , \qquad
\bt_3^*(\al_s) \; \simeq \; - 0.362 \al_s^{2.603} \qquad ( \al_s \ra \infty) \;  .
\ee

\section{Discussion}

Summarizing, self-similar approximation theory, presented in the report, is a powerful 
and at the same time rather simple tool for extrapolating asymptotic series from the 
region of small values of a parameter or variable to their finite values and even to the 
large values tending to infinity. Because the limiting case of large variables is often
of the most interest and also it is the most difficult to study, we paid more attention 
to this problem. The limiting case of large variables can be obtained knowing solely 
expansions in powers of asymptotically small variables, without any information on the 
large-variable limit, except assuming that it is of power law. If, in addition to the 
small-variable expansion, there exists a large-variable expansion, then the self-similar 
approximation theory \cite{Yukalov_10,Yukalov_11} allows us to construct approximants 
interpolating between weak-coupling and strong-coupling expansions \cite{Yukalov_54}. 

In Sec. 2, we presented an outline of the basic steps of self-similar approximation 
theory, allowing the reader to grasp why this theory makes it possible to reconstruct
the whole functions from the knowledge of their weak-coupling or small-variable 
expansions. To make the long story short, this is because the terms of an asymptotic 
expansion, although being derived in one region of the variable, say in the region of 
small variables, contain information on the whole function they have been obtained 
from. This information is hidden in the structure of the approximate-term variation
between subsequent approximation orders. The theory deciphers this hidden structure by
extracting the self-similar properties in the series of the subsequent approximation terms. 
Such a deciphering becomes possible by representing the sequence of approximation terms 
as a trajectory of a dynamical system in discrete time played by the approximation order.

To give more understanding on the properties of self-similar approximants, their 
convergence, and the domain of applicability, let us recall the related useful theorems. 
We keep in mind real functions of a real variable $x \in [0,\infty)$.  

\vskip 2mm

{\bf Theorem 1} \cite{Yukalov}. If the sequence $\{f_k(x)\}$ of the terms
\be
\label{31}
f_k(x) \; = \; \sum_{n=0}^k a_n x^n \;   ,
\ee
with $a_0 = 1$, converges on a domain $\mathbb{D} \subset \mathbb{R}_+$ to a function    
\be
\label{32}
f(x) \; = \; \sum_{n=0}^\infty a_n x^n \;    ,
\ee
then the sequence $\{f^*_k(x)\}$ of the terms
\be
\label{33}
f_k^*(x) \; = \; \prod_{j=1}^{N_k} ( 1 + A_j x )^{n_j} \;   ,
\ee
with the control parameters $A_j$ and $n_j$ satisfying the conditions
\be
\label{34}
\lim_{x\ra 0} \; \frac{1}{n!} \; \frac{d^n}{dx^n} \; f_k^*(x) \; = \; a_n \;   ,
\ee
converges on this domain to the function
\be
\label{35}
f^*(x) \; = \; \prod_{j=1}^\infty ( 1 + A_j x )^{n_j} \; 
\ee
coinciding with $f(x)$,
\be
\label{36}
 f^*(x) \; = \; f(x) \;  .
\ee

\vskip 2mm

{\it Comment 1}. This theorem tells us that the method of self-similar factor 
approximants is regular.

\vskip 2mm
{\bf Theorem 2} \cite{Yukalov_20}. The sequence of the terms $\{f^*_k(x)\}$ converges
if and only if the sequence of the sums
\be
\label{37}
S_k(x) \; = \; \sum_{j=1}^{N_k} n_j \ln ( 1 + A_j x)
\ee
converges.

\vskip 2mm

{\it Comment 2}. If the sequence of the sums (\ref{37}) converges for a fixed 
$x \in \mathbb{D}$, then  
\be
\label{38}
 \lim_{j\ra\infty} n_j A_j \; = \; 0 \;  .
\ee

\vskip 2mm
{\bf Theorem 3} \cite{Yukalov_20}. If $A_j x \geq -1$ and the sequence of the sums
\be
\label{39}
s_k \; = \; \sum_{j=1}^{N_k} |\; n_j A_j \; |
\ee
converges, then the sequence of the apporximants $\{f^*_k(x)\}$ converges.

\vskip 2mm
{\it Comment 3}. If the sequence $\{s_k\}$ converges, then the scaling 
\be
\label{40}
A_j \; \longmapsto \; \lbd_j A_j \; , \qquad n_j \; \longmapsto \; \frac{n_j}{\lbd_j}
\ee
does not change the convergence of $\{s_k\}$.

\vskip 2mm
{\bf Theorem 4} \cite{Yukalov_17,Gluzman_18,Yukalov_47}. Let $F_N(x)$ be a real function 
of a real variable $x$, having the form
\be
\label{41}
 F_N(x) \; = \; \prod_{j=1}^N P_{m_j}^{\al_j} (x) \;  ,
\ee
where $N \geq 2$ is an integer, $P_m(x)$ are polynomials of degree $m$ and $\alpha_j$ 
are either real numbers or complex-valued numbers entering by complex-conjugate pairs,
so that $F_N(x)$ is real. Then the function $F_N(x)$ is exactly reproducible by the 
factor approximants of order
\be
\label{42}
 k \; \geq \; \sum_{j=1}^N m_j \;  .
\ee
    
\vskip 2mm
{\it Comment 4}. In addition to the class of functions (\ref{41}), the exponential 
function $f(x) = e^x$ is also exactly reproducible from its finite-order expansion
$f_k(x) = \sum_{n=0}^k x^n/n!$ by the factor approximants of order $k \geq 2$, 
which follows directly from condition (\ref{34}), 
$$
 f_2^*(x) \; = \; \lim_{A\ra 0} ( 1 + A x)^{1/A} \; = \; e^x \; , 
$$
$$
f_3^*(x) \; = \; \lim_{A\ra 0} ( 1 + x)^{A/(1-A)} \; ( 1 + A x)^{1/A(1-A)} \; =
 \; e^x \;  ,
$$  
and so on. 

\vskip 2mm
These theorems, in each particular case, allow us to understand when and why asymptotic 
series can be effectively summed by employing self-similar approximation theory. In 
practice, however, one usually deals not with infinite series, but only with finite 
expansions derived by means of some kind of perturbation theory. In that case, one has 
to resort to the standard method of analysis of obtained numerical data by comparing the 
available approximation terms and observing wether there exists numerical convergence. 
As an example, let us consider the case of the ground-state energy of one-dimensional 
anharmonic oscillator, for which a finite expansion in powers of the coupling parameter 
$g \ra 0$ is known \cite{Bender_55,Hioe_56}. There are several methods of extrapolating 
the weak-coupling expansions to finite values of the parameters $g$, for instance Pad\'{e} 
approximants. But such methods cannot solve the most difficult problem of extrapolating 
a weak-coupling expansion to the strong-coupling limit $g \ra \infty$. Applying to the 
weak-coupling expansion the method of self-similar Borel summation, we find 
\cite{Yukalov_20} the strong-coupling limit of the ground-state energy 
\be
\label{43}     
 E_k^*(g) \; \simeq \; C_k g^{\nu_k} \;  .
\ee
The strong-coupling amplitude $C_k$ and exponent $\nu_k$, for different approximation orders 
$k$, are listed in Table 1. These numerical data show that the amplitude converges to the 
value $C_{14} = 0.688$, while the exponent to the value $\nu_{14} = 0.327$, which are 
very close to the exact numerical data $C = 0.688$ and $\nu = 1/3$.   

\begin{table}[hp]
%Table 1
\centering
\renewcommand{\arraystretch}{1.2}
\begin{tabular}{|l|c|c|} \hline
$k$   &  $C_k$  &  $\nu_k$      \\ \hline
2     &  0.727  &    0.300  \\ 
3     &  0.727  &    0.289   \\ 
4     &  0.727  &    0.289  \\ 
5     &  0.713  &    0.310    \\
6     &  0.712  &    0.312    \\ 
7     &  0.702  &    0.319    \\
10    &  0.698  &    0.322    \\ 
11    &  0.695  &    0.324   \\ 
13    &  0.690  &    0.326   \\ 
14    &  0.688  &    0.327  \\ \hline
exact &  0.668  &    0.333 \\ \hline
\end{tabular}
\caption{\small
Strong-coupling amplitudes and exponents for the ground-state energy of one-dimensional
anharmonic oscillator, predicted by self-similar Borel summation.}
\end{table}

In conclusion, when one needs to extrapolate a small-variable expansion to finite and even 
large values of the variable, the straightforward recipe is: (i) Apply the self-similar 
approximation techniques either directly or by involving self-similar Borel summation.
(ii) Choose the sequence of the approximants that exhibits a better numerical convergence.
(iii) If there are no real-valued approximants, or there is no numerical convergence, check 
by resorting to the above theorems, where could be the problem. Usually, the absence of
convergence means that the available number of terms in the studied expansion is not 
sufficient for extracting the self-similar scaling hidden in the given expansion terms, 
and more terms are required. In some cases, numerical convergence can be achieved by
introducing, in the frame of the self-similar scheme, additional control parameters and 
using more refined Borel-type transformations, as is shown in \cite{Yukalov_57}.

\vskip 1cm
{\parindent=0pt

\vskip 5mm
{\bf Conflicts of interest}

\vskip 2mm
The authors of this work declare that they have no conflicts of interest.

}

\end{document}